\documentclass[prd,reprint,nofootinbib,preprintnumbers,showpacs]{revtex4-1}
\usepackage{hyperref,amsmath,amssymb,graphicx,bm,bbm}
\newcommand{\arXivid}[1]{\href{http://arxiv.org/abs/#1}{arXiv:#1}}

\begin{document}
\preprint{PUPT-2412}
\title{Analytic fermionic Green's functions from holography}
\author{Steven S. Gubser}
\author{Jie Ren}
\affiliation{Department of Physics, Princeton University, Princeton, NJ 08544}
\date{\today}

\def\SSG#1{{\bf [SSG: #1]}}

\begin{abstract}
  We find exact, analytic solutions of the Dirac equation for a charged, massless fermion in the background of a charged, dilatonic black hole in AdS$_5$. The black hole descends from type IIB supergravity, where it describes D3-branes with equal angular momenta in two of the three independent planes of rotation orthogonal to the world-volume. The Green's function near the Fermi surface for a strongly coupled fermionic system can be extracted holographically from an exact solution of the Dirac equation at zero frequency but nonzero momentum. There can be several Fermi momenta, and they take the form $k_F=q-n-1/2$ (in units of the chemical potential), where $q$ is the charge of the spinor, and $n$ is a non-negative integer that labels the Fermi surfaces. Much as for holographic Fermi surfaces based on the Reissner-Nordstr\"{o}m-AdS$_5$ solution, the dispersion relation of the excitations near the Fermi surface is determined by the geometry close to the horizon, and one can obtain Fermi liquid, marginal Fermi liquid, and non-Fermi liquid behaviors depending on the value of $k_F$.
\end{abstract}
\maketitle

\section{Introduction}
Charged black holes in asymptotically anti-de Sittter (AdS) spacetime can be regarded as the gravitational dual description of certain strongly interacting fermionic systems at finite charge density, such as non-Fermi liquids \cite{Lee:2008xf}. This is an application of the gauge/gravity duality \cite{Maldacena:1997re,Gubser:1998bc,Witten:1998qj}, which allows us to calculate fermionic Green's function by solving the bulk Dirac equation \cite{Henningson:1998cd,Mueck:1998iz,Iqbal:2009fd}.  A particularly well-studied example is the Reissner-Nordstr\"{o}m (RN) black hole in AdS as the geometry \cite{Liu:2009dm,Cubrovic:2009ye,Faulkner:2009wj,Faulkner:2011tm}.  We will focus instead on a particular dilatonic black hole in AdS$_5$, explored originally in Ref.~\cite{Gubser:2009qt}; see also Ref.~\cite{Goldstein:2009cv} for related work.

The dilatonic black hole in question is sometimes referred to as the two-charge black hole.  From a five-dimensional point of view, this is because two of the three mutually commuting $U(1)$ subgroups of the $SO(6)$ gauge group of maximal gauged supergravity are nonzero and equal, while the third is zero.  From a ten-dimensional point of view, this black hole describes $N$ coincident D3-branes with equal, nonzero angular momentum in two of the three independent planes of rotation orthogonal to the D3-brane world-volume.  The dilatonic black hole enjoys several advantages over the better studied RN-AdS$_5$ black hole:
\begin{itemize}
  \item The entropy and specific heat of the dilatonic black hole are proportional to temperature at low temperature, as compared to a nonzero, ${\cal O}(N^2)$ entropy at extremality for the RN-AdS$_5$ black hole.
  \item Exact information about the position and properties of Fermi surfaces is available for the dilatonic black hole, for massless bulk fermion actions with no Pauli couplings.  This stands in contrast with the RN-AdS$_5$ black hole, where one must resort to numerics to find $k_F$.  (This is even true of Ref.~\cite{DeWolfe:2011aa}, in which numerical work led to strong evidence that the Fermi momenta are simple algebraic numbers.)
  \item Pair creation of fermions near the horizon, and back-reaction of the resulting fermionic matter, must distort the RN-AdS$_5$ geometry to some extent.  But for the dilatonic black holes there is some evidence, to be explained below, that pair creation of fermions is suppressed.
\end{itemize}
A notable disadvantage of the dilatonic black hole is that its extremal limit---which will be our main focus---has a naked singularity.  Any nonzero temperature cloaks the naked singularity with a horizon, but as temperature is taken to zero, the dilaton as well as curvature invariants become larger and larger at the horizon, until at zero temperature they diverge.  Nevertheless it is straightforward to pick out physically reasonable boundary conditions for fermions: In particular, for $\omega=0$ one can simply demand that the allowed solutions are regular as the naked singularity is approached.

The main aim of the current work is to solve the massless Dirac equation,
\begin{equation}
\gamma^\mu(\nabla_\mu-iqA_\mu)\Psi=0, \label{DiracEq}
\end{equation}
in the extremal limit of the dilatonic black hole background, and to show that the corresponding Green's function exhibits one or more Fermi surfaces if $q > 1/2$.  For $1/2 < q < 1$, there is only a single Fermi surface, and $v_F$ is not well-defined.  For $1 < q < 3/2$, there is still only a single Fermi surface, but $v_F$ is well defined.  For $q > 3/2$, there are additional Fermi surfaces at $k_F = q - n - 1/2$, where $n$ is a positive integer.  The outermost Fermi surface has the simplest properties: assuming $q>1$, the Green's function near the Fermi surface takes the form
\begin{equation}
G=\frac{Z}{-\omega+v_F(k-k_F)-\Sigma(\omega,k_F)},
\end{equation}
where
\begin{gather}
k_F=q-\frac{1}{2}\qquad v_F=\frac{4(q-1)}{4q-3} \nonumber \\
\Sigma=\frac{\Gamma(q+1/2)\Gamma(1-q)e^{i\pi(1-q)}}{2^{4q-5}\sqrt{\pi}(4q-3)\Gamma(q-1)\Gamma(q)}\omega^{2q-1} \nonumber \\
Z=\frac{8\Gamma(q+1/2)}{\sqrt{\pi}(4q-3)\Gamma(q-1)}.  \label{eq:vSigmaZ}
\end{gather}
Formulas generalizing Eq.~\eqref{eq:vSigmaZ} to Fermi surfaces with $n>0$ can be found in Sec.~\ref{sec:green}.

The organization of the rest of this paper is as follows.  In Sec.~\ref{sec:nm}, we solve the Dirac equation at $\omega=0$ in terms of hypergeometric functions, and find the normal modes that determine the location of Fermi surfaces.  In Sec.~\ref{sec:IR}, we study the near horizon geometry (hereafter IR for infrared), solving the Dirac equation and obtaining the IR Green's function. In Sec.~\ref{sec:green}, we obtain the Green's function near Fermi surface by matching the IR solution to a zero-frequency solution away from the IR.  In Sec.~\ref{sec:full}, we numerically solve the Green's function and explain the main features at general $\omega$. In Sec. \ref{sec:sum}, we conclude with some discussions.

\section{Normal modes}\label{sec:nm}
The two-charge black hole in AdS$_5$ is determined by
\begin{align}\label{eq:StartingLagrangian}
{\mathcal L}=&\frac{1}{2\kappa^2}\Bigl[R-\frac{1}{4}e^{4\alpha}F_{\mu\nu}^2\notag\\
&-12(\partial_\mu\alpha)^2+\frac{1}{L^2}(8e^{2\alpha}+4e^{-4\alpha})\Bigr],
\end{align}
which is from a consistent truncation of the type IIB supergravity with three U(1) charges $Q_1=Q_2=Q$ and $Q_3=0$. The solution in the extremal case is
\begin{gather}
ds^2=e^{2A}(-hdt^2+d{\bf x}^2)+\frac{e^{2B}}{h}dr^2\nonumber\\
A=\ln\frac{r}{L}+\frac{1}{3}\ln\left(1+\frac{Q^2}{r^2}\right)\nonumber\\
B=-\ln\frac{r}{L}-\frac{2}{3}\ln\left(1+\frac{Q^2}{r^2}\right)\nonumber\\
h=\frac{(r^2+2Q^2)r^2}{(r^2+Q^2)^2}\qquad \alpha=\frac{1}{6}\ln\left(1+\frac{Q^2}{r^2}\right)\nonumber\\
A_\mu dx^\mu=\Phi dt\qquad \Phi=\frac{\sqrt{2}Qr^2}{(r^2+Q^2)L}.
\end{gather}
The ``horizon" for this black hole is at $r=0$, which is a spacetime singularity. For the non-extremal case, and its ten-dimensional lift, see Ref.~\cite{Gubser:2009qt}.

We will solve the Dirac equation for a massless spinor in the above background, but we keep the mass term at first. If the metric is diagonal and depends only on the radial coordinate $r$, the Dirac equation can be simplified by using the rescaled spinor $\tilde{\Psi}=(-gg^{rr})^{1/4}\Psi$. The equation of motion for $\tilde{\Psi}$ is
\begin{equation}
[\gamma^\mu(\partial_\mu-iqA_\mu)-m]\tilde{\Psi}=0.\label{eq:tilpsi}
\end{equation}
We assume that the momentum is in the $x$ direction. By plugging a single Fourier mode $\tilde{\Psi}\sim e^{-i\omega t+ikx}\hat{\Psi}$ to Eq.~\eqref{eq:tilpsi}, the equation for $\hat{\Psi}$ is
\begin{multline}
\bigl[-i\sqrt{-g^{tt}}\gamma^{\underline{t}}(\omega+qA_t)+\sqrt{g^{rr}}\gamma^{\underline{r}}\partial_r\\
+i\sqrt{g^{xx}}\gamma^{\underline{x}}k-m\bigr]\hat{\Psi}=0.\label{eq:dirac}
\end{multline}
We choose the following gamma matrices for AdS$_5$:
\begin{gather}
\gamma^{\underline{t}}=\begin{pmatrix}
i\sigma_1 & 0\\
0 & i\sigma_1
\end{pmatrix}\qquad
\gamma^{\underline{r}}=\begin{pmatrix}
\sigma_3 & 0\\
0 & \sigma_3
\end{pmatrix}\notag\\
\gamma^{\underline{x}}=\begin{pmatrix}
-\sigma_2 & 0\\
0 & \sigma_2
\end{pmatrix}\qquad
\gamma^{\underline{y}}=\begin{pmatrix}
0 & -\sigma_2\\
-\sigma_2 & 0
\end{pmatrix}\notag\\
\gamma^{\underline{z}}=\begin{pmatrix}
0 & i\sigma_2\\
-i\sigma_2 & 0
\end{pmatrix}.
\end{gather}
Then Eq.~\eqref{eq:dirac} reduces to two decoupled equations
\begin{multline}
\bigl[\sqrt{-g^{tt}}\sigma_1(\omega+qA_t)+\sqrt{g^{rr}}\sigma_3\partial_r\\
+(-1)^\alpha\sqrt{g^{xx}}i\sigma_2k-m\bigr]\psi_\alpha=0,\label{eq:psi12}
\end{multline}
where $\psi_1$ and $\psi_2$ are two-component spinors. The equation for $\psi_2$ is related to the equation for $\psi_1$ by $k\to-k$.

The (massive or massless) Dirac spinor $\Psi$ in the AdS$_5$ maps to a chiral spinorial operator $\mathcal{O}_\Psi$ at the boundary \cite{Henningson:1998cd,Mueck:1998iz,Iqbal:2009fd}. The asymptotic behavior of $\psi_\alpha$ near the AdS boundary is
\begin{equation}
\psi_\alpha
\xrightarrow{r\to\infty} a_\alpha r^{m}\begin{pmatrix}
1\\
0
\end{pmatrix}
+b_\alpha r^{-m}\begin{pmatrix}
0\\
1
\end{pmatrix}.
\end{equation}
The expectation value of the boundary spinorial operator dual to the bulk spinor $\Psi$ has the form $\langle \mathcal{O}_\Psi \rangle=(0,b_1,0,b_2)^T$. In fact, $\mathcal{O}_\Psi=\frac{1}{2}(1-\gamma^{\underline{r}})\mathcal{O}_\Psi$, which means that the boundary spinorial operator is left-handed. By imposing the in-falling boundary condition at the horizon, we can obtain the retarded Green's function as
\begin{equation}
G=\begin{pmatrix}
0 & & &\\
& G_1 & &\\
& & 0 &\\
& & & G_2
\end{pmatrix},\qquad G_\alpha=\frac{b_\alpha}{a_\alpha}.\label{eq:G12}
\end{equation}

Note that if we use the alternative quantization, the Green's function is $\tilde{G}_\alpha=-a_\alpha/b_\alpha$, and the boundary spinorial operator is right-handed. If $m=0$, $G_1$ and $G_2$ are related by $G_2=-1/G_1$ \cite{Faulkner:2009wj}; therefore, the alternative quantization for $G_1$ is the standard quantization for $G_2$, and vice versa. By taking into account both $G_1$ and $G_2$, the alternative quantization gives the same Fermi momenta as the standard quantization does, when $m=0$ \cite{Faulkner:2009wj}.

We will focus on $\psi_1\equiv (u_1,u_2)^T$ in the following. The square roots in the Dirac equation can be eliminated, following a method which has appeared, for example, in Ref.~\cite{Batic:2006hy}. Define $u_\pm=u_1\pm iu_2$. From Eq.~\eqref{eq:psi12}, we obtain
\begin{align}
u_+'+\bar{\lambda}(r)u_+ &=\bar{f}(r)u_-\label{eq:uplus}\\
u_-'+\lambda(r)u_- &=f(r)u_+,\label{eq:uminus}
\end{align}
where
\begin{equation}
\lambda(r)=i\sqrt{\frac{|g^{tt}|}{g^{rr}}}(\omega+qA_t),\qquad
f(r)=\frac{m}{\sqrt{g^{rr}}}-ik\sqrt{\frac{g^{xx}}{g^{rr}}}.
\end{equation}
The Eqs.~\eqref{eq:uplus} and \eqref{eq:uminus} can be decoupled to obtain two second-order differential equations:
\begin{align}
u_+''+\bar{p}(r)u_+'+\bar{q}(r)u_+ &=0\label{eq:up}\\
u_-''+p(r)u_-'+q(r)u_- &=0,\label{eq:um}
\end{align}
where
\begin{equation}
p(r)=-\frac{f'}{f},\qquad q(r)=|\lambda|^2-|f|^2+p\lambda+\lambda'.
\end{equation}
After we solve Eq.~\eqref{eq:um} for $u_-$, we need to plug in $u_-$ to Eq.~\eqref{eq:uminus} to obtain $u_+$.\footnote{We cannot only solve Eqs.~\eqref{eq:up} and \eqref{eq:um} and discard Eqs.~\eqref{eq:uplus} and \eqref{eq:uminus}, because there are only two boundary conditions. After we solve $u_-$ from the second-order equation (\ref{eq:um}), $u_+$ is fully determined by the first-order equation \eqref{eq:uminus}.} For the metric we consider,
\begin{equation}
\lambda=\frac{i(\omega+q\Phi)}{he^{A-B}},\qquad f=\frac{m}{\sqrt{h}e^{-B}}-\frac{ik}{\sqrt{h}e^{A-B}}.
\end{equation}

In the following, we will only study the $m=0$ case, in which $p(r)$ and $q(r)$ are rational functions of $r$. We are most interested in the following two questions: whether there are Fermi surfaces, and whether there are quasiparticles near the Fermi surfaces. We will solve the Dirac equation at $\omega=0$ first, and the solution indicates that there are one or more Fermi surfaces when $q>1/2$, as summarized more precisely in the text following Eq.~(\ref{DiracEq}). Then the perturbation at small $\omega$ will give the Green's function near the Fermi surfaces.

When $\omega=0$, the boundary condition for at the horizon is that the solution is regular. The solution for $u_\pm$ can be written as\footnote{Note that $(-1)^\alpha:=(-1+i\epsilon)^\alpha=e^{i\pi\alpha}$ and $(-1-i\epsilon)^\alpha=e^{-i\pi\alpha}$.}
\begin{multline}
u_-=\left(\frac{r}{r+i\sqrt{2}Q}\right)^{\nu_k}\left(\frac{r+i\sqrt{2}Q}{r-i\sqrt{2}Q}\right)^{q/2}\\
\times{_2F_1}\left(\nu_k-q+\frac{1}{2},\,\nu_k;\,2\nu_k+1;\,\frac{2r}{r+i\sqrt{2}Q}\right)\label{eq:umsol}
\end{multline}
and
\begin{equation}
u_+=(-1)^{-\nu_k+q+1/2}u_-^*,\label{eq:upsol}
\end{equation}
where
\begin{equation}
\nu_k=\frac{k}{\sqrt{2}Q}.
\end{equation}
The chemical potential $\sqrt{2}Q$ is a unit of the energy scale. To have physical bound states, $Q$ and $q$ must have the same sign; we assume $Q>0$ and $q>0$. This system has rotational invariance; we can choose $\textbf{k}=(k,0,0)$, where $k>0$. Thus we have $\nu_k>0$, without loss of generality.

By defining $\nu_k-q+1/2=-n$, the solution for $u_1$ and $u_2$ is
\begin{align}
u_1 &=\frac{u_++u_-}{2}=\frac{(-1)^{n+1}u_-^*+u_-}{2}\label{eq:u1sol}\\
u_2 &=\frac{u_+-u_-}{2i}=\frac{(-1)^{n+1}u_-^*-u_-}{2i}.\label{eq:u2sol}
\end{align}
The Green's function $G(\omega,k)$ at $\omega=0$ is real:
\begin{equation}
G_1=\lim_{r\to\infty}\frac{u_2}{u_1}
=\lim_{r\to\infty}\left(-i\frac{(-1)^{n+1}u_-^*-u_-}{(-1)^{n+1}u_-^*+u_-}\right)=G_1^*.
\end{equation}
This apparently implies that the spectral density is zero at $\omega=0$. However, we need to shift the pole at $\omega=0$ by $\omega\to\omega+i\epsilon$, and then we will obtain a delta function in the imaginary part.\footnote{For example, for a free electron near $k_F$ ($k_\perp\equiv k-k_F$): $$\frac{1}{-\omega+v_Fk_\perp-i\epsilon}
=\mathcal{P}\frac{1}{-\omega+v_Fk_\perp}+i\pi\delta(\omega-v_Fk_\perp).$$}

\begin{figure}
\centering
\includegraphics[]{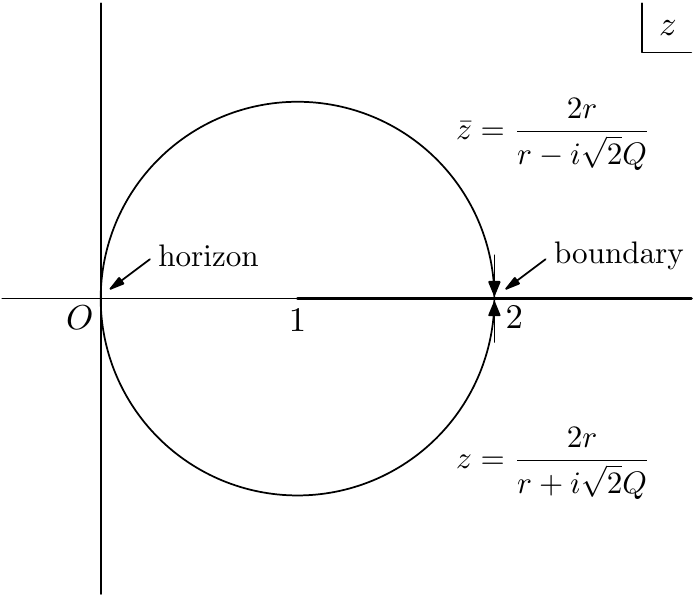}
\caption{\label{fig:zplane} The real axis in the complex $r$-plane maps to a circle in the complex $z$-plane. The hypergeometric function $F(\alpha,\beta;\gamma;z)$ has a branch cut from $z=1$ to $\infty$ in general, but the branch cut is absent when $\alpha$ is a non-negative integer.}
\end{figure}

The normal modes are determined by $u_1|_{r\to\infty}=0$. In general, the hypergeometric function ${_2F_1}(\alpha,\beta;\gamma;z)$ has a branch cut from $z=1$ to $\infty$. At the AdS boundary,
\begin{equation}
u_-|_{r\to\infty}={_2F_1}(-n,\nu_k;2\nu_k+1;2-i\epsilon).
\end{equation}
Thus, $u_-$ and $u_-^*$ take values at different sides of the branch cut, as shown in Fig.~\ref{fig:zplane}. However, if $\alpha=-n$, where $n=0$, $1$, $2$, $\cdots$, the hypergeometric function is an $n$th-order polynomial of $z$, and the branch cut from $z=1$ to $\infty$ is absent. More generally, the equation that determines the normal modes is Eq.~\eqref{eq:nmodes} in appendix \ref{sec:notes}, in which we conclude that there are no physical solutions when $n$ is not a non-negative integer.

At the AdS boundary $r\to\infty$, $u_-^*=u_-$ if $\alpha=-n$, where $n=0$, $1$, $2$, $\cdots$. Therefore, if $n$ is even, $\nu_k^{(n)}=q-n-1/2$ gives the Fermi surface for the standard quantization ($u_1=0$); if $n$ is odd, $\nu_k^{(n)}=q-n-1/2$ gives the Fermi surface for the alternative quantization ($u_2=0$). This conclusion is for the Green's function $G_1$, which obtained by the upper-half components of the bulk spinor. Recall that $G_2=-1/G_1$, which is obtained by the lower-half components of the bulk spinor. In the following, we use the standard quantization only. Taking into account both $G_1$ and $G_2$, we conclude that the Fermi momenta are determined by $\nu_k^{(n)}=q-n-1/2$, where $n$ is a non-negative integer such that $q-n-1/2>0$. Note that the alternative quantization gives the same Fermi momenta, with the difference that the boundary fermionic operator is right-handed.

By perturbation, we can obtain the analytic solution of the Green's function near the Fermi surface. The Green's function can be written as
\begin{equation}
G_R(\omega,k)=\frac{Z}{-\omega+v_F(k-k_F)-\Sigma(\omega,k_F)},\label{eq:green}
\end{equation}
where $k_F=|\mathbf{k}_F|$ is the Fermi momentum, $v_F$ is the Fermi velocity, and
\begin{equation}
\Sigma(\omega,k)=h(k){\cal G}_k(\omega),\qquad {\cal G}_k(\omega)=c(k)\omega^{2\nu_k}.
\end{equation}
As fermionic Green's functions, $G$ and $\mathcal{G}$ satisfy $\text{Im}(G)>0$ and $\text{Im}(\mathcal{G})>0$ for all real $\omega$. The result shows that $v_F>0$, $Z>0$, and $h>0$. The Fermi momenta are determined by
\begin{equation}
\frac{k_F^{(n)}}{\sqrt{2}Q}=q-n-\frac{1}{2},
\end{equation}
where $n=0$, $1$, $2$, $\cdots$, $\lfloor q-1/2\rfloor$. When $n$ is even, Eq.~\eqref{eq:green} is for $G_1$; when $n$ is odd, Eq.~\eqref{eq:green} is for $G_2$. Again note that we need to shift the $\omega=0$ pole to the lower half complex $\omega$-plane by $\omega\to\omega+i\epsilon$ to obtain a well-defined retarded Green's function.

\section{IR geometry and Green's function}\label{sec:IR}
We expect that the Green's function near the Fermi surface can be obtained by the perturbation of small $\omega$ around the exact solution. However, the ordinary perturbation method is not enough when the black hole is extremal. As pointed out in Ref.~\cite{Faulkner:2009wj}, when it is sufficiently close to the horizon, $\omega$-dependent terms cannot be treated as small perturbations no matter how small $\omega$ is. This section and the next are in parallel with Ref.~\cite{Faulkner:2009wj}, in which a systematic method is developed for treating the extremal black hole system. Usually this method relies on numerics to fix certain quantities, such as the Fermi velocity. The example we provide is exactly solvable, in the sense that a perturbative treatment of the small $\omega$ regime can be obtained through matched asymptotic expansions of analytically known functions.

\begin{figure}
\includegraphics[]{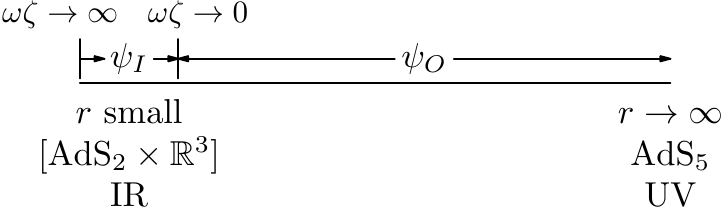}
\caption{\label{fig:match} The inner (near horizon) and outer regions, where the solutions of the Dirac equation are denoted by $\psi_I$ and $\psi_O$, respectively.}
\end{figure}

We divide the geometry into inner and outer regions, as shown in Fig.~\ref{fig:match}. The inner region refers to the IR (near horizon) geometry, in which the Dirac equation can be exactly solved to give an IR Green's function. The outer region refers to the remaining geometry, in which we can make perturbations for small $\omega$. Then we need to match the inner and outer regions.

The IR geometry is examined as follows. In the $r\to 0$ limit, the metric becomes
\begin{equation}
ds^2=\left(\frac{r}{Q}\right)^{2/3}\left(-\frac{2r^2}{L^2}dt^2+\frac{L^2}{2r^2}dr^2+\frac{Q^2}{L^2}d{\bf x}^2\right).\label{eq:ads2}
\end{equation}
Therefore, the IR geometry is conformal to $\text{AdS}_2\times\mathbb{R}^3$. This can be made more explicit by change of variables
\begin{equation}
r=\frac{L_2^2}{\zeta},\qquad L_2=\frac{L}{\sqrt{2}},
\end{equation}
and the metric becomes
\begin{equation}
ds^2=\left(\frac{L^2}{2Q\zeta}\right)^{2/3}\left[\frac{L_2^2}{\zeta^2}\left(-dt^2+d\zeta^2\right)+\frac{Q^2}{L^2}d{\bf x}^2\right].
\end{equation}
The gauge field $A_t$ becomes
\begin{equation}
\Phi=\frac{L_2^3}{Q\zeta^2}.
\end{equation}
We will switch back to the $r$ coordinate. Note that in the RN-AdS black hole system, $\Phi\sim r$, and thus the electric field $E=\nabla\Phi$ is constant at the horizon. In our system, $\Phi\sim r^2$, and thus the electric field $E=\nabla\Phi\sim r$ falls off toward the horizon. This leads to a significant difference relative to the Dirac equation in AdS$_2$. In the near horizon limit $r\to 0$ ($\zeta\to\infty$), the contribution by the electric field to the Dirac equation is negligible. Nevertheless, the flux is conserved by $d(e^{4\alpha}\,{^*\! F})=0$.

We solve the Dirac equation in the geometry Eq.~\eqref{eq:ads2} without the electric field. The solution for $u_\pm$ with in-falling boundary condition\footnote{The in-falling wave in terms of the coordinate $\zeta$ is $e^{i\omega\zeta}$ as $\zeta\to\infty$.} is
\begin{align}
u_- &=C\sqrt{r}\,W_{1/2,\nu_k}(-i\omega/r)\label{eq:uIplus}\\
u_+ &=i\nu C\sqrt{r}\,W_{-1/2,\nu_k}(-i\omega/r),\label{eq:uIminus}
\end{align}
where $W$ is a Whittaker function, and $C$ is a constant. Denote $\psi_I$ as the solution in the inner region. In the near boundary limit of the IR geometry, the asymptotic behavior is
\begin{eqnarray}
\psi_I &\to& \alpha\left(\frac{\omega}{r}\right)^{-\nu_k}+\beta\left(\frac{\omega}{r}\right)^{\nu_k}
\qquad{\rm as}\quad\frac{\omega}{r}\to 0.
\end{eqnarray}
More precisely, the inner region solution can be written as
\begin{equation}
\psi_I=v_+r^{\nu_k}(1+\cdots)+{\cal G}_k(\omega)v_-r^{-\nu_k}(1+\cdots),\label{eq:psiI}
\end{equation}
where $v_\pm$ must be chosen to match the normalization of Eqs.~\eqref{eq:umsol} and \eqref{eq:upsol}. By expanding $\psi_I=(u_1,u_2)^T$ from Eqs.~\eqref{eq:uIplus} and \eqref{eq:uIminus}, we know that $v_\pm$ take the following form
\begin{equation}
v_+=\lambda_+\begin{pmatrix}
1\\
1
\end{pmatrix},\qquad
v_-=\lambda_-\begin{pmatrix}
-1\\
1
\end{pmatrix},
\end{equation}
where $\lambda_\pm$ are constants. The Green's function $\mathcal{G}_k(\omega)$ depends on the ratio $\lambda_+/\lambda_-$. However, the self-energy $\Sigma$ is independent of $\lambda_\pm$ after matching the inner and outer regions. We choose $\lambda_+=\lambda_-$, and then the IR Green's function is\footnote{Other ways to write down $\mathcal{G}_k(\omega)$ are $$-ie^{-i\pi\nu}\frac{\Gamma(-2\nu)\Gamma(1+\nu)}{\Gamma(2\nu)\Gamma(1-\nu)}\omega^{2\nu},\qquad \frac{(\tan\pi\nu+i)\pi}{\Gamma(\nu+1/2)^2}\left(\frac{\omega}{4}\right)^{2\nu}.$$}
\begin{equation}
{\cal G}_k(\omega)=e^{i\pi(1/2-\nu_k)}\frac{\Gamma(1/2-\nu_k)}{\Gamma(1/2+\nu_k)}\left(\frac{\omega}{4}\right)^{2\nu_k}.
\end{equation}
The IR Green's function can be generalized to finite temperature when the back hole is near extremal:
\begin{equation}
{\cal G}_k(\omega)=i\Bigl(\frac{\pi T}{2}\Bigr)^{2\nu_k}\frac{\Gamma\bigl(\frac{1}{2}-\nu_k\bigr)\Gamma\bigl(\frac{1}{2}+\nu_k-\frac{i\omega}{2\pi T}\bigr)}{\Gamma\bigl(\frac{1}{2}+\nu_k\bigr)\Gamma\bigl(\frac{1}{2}-\nu_k-\frac{i\omega}{2\pi T}\bigr)}.
\end{equation}

The main difference between the RN-AdS black hole system and our system is attributed to the IR geometry with the gauge field. In the RN-AdS$_5$ black hole system, the IR geometry is $\text{AdS}_2\times\mathbb{R}^3$, and the electric field is nonzero at the horizon. The IR scaling exponent has the form $\nu_k=\sqrt{k^2-k_o^2}$, which depends on the charge of the spinor, and will become imaginary if the charge is large. The system with this IR behavior is studied as a semi-local quantum liquid \cite{Iqbal:2011in}. The imaginary $\nu_k$ implies an instability causing by the pair production near the black hole horizon \cite{Faulkner:2009wj,Pioline:2005pf}. It has been argued that back-reaction from the pair production alters the IR region to a Lifshitz geometry \cite{Hartnoll:2009ns}; a candidate of the final geometry was constructed as the electron star \cite{Hartnoll:2010gu,Hartnoll:2010ik,Hartnoll:2011dm}. In our system, $\nu_k$ is always real, and the electric field approaches zero in the near horizon limit.

\section{Green's function near the Fermi surface}\label{sec:green}
In the outer region, the solution at small $\omega$ can be written as
\begin{equation}
\psi_O=\eta_++{\cal G}_k(\omega)\eta_-,
\end{equation}
where
\begin{equation}
\eta_\pm=\eta_\pm^{(0)}+\omega\eta_\pm^{(1)}+\omega^2\eta_\pm^{(2)}+\cdots.
\end{equation}
The asymptotic behavior near the horizon is
\begin{equation}
\eta_\pm^{(0)}=v_\pm r^{\pm\nu_k}+\cdots,\qquad r\to 0,
\end{equation}
which is matched with the inner region solution, Eq.~\eqref{eq:psiI}. Here $\eta_+^{(0)}=(u_1,u_2)^T$, where $u_1$ and $u_2$ are solutions in the outer region as Eqs.~\eqref{eq:u1sol} and \eqref{eq:u2sol}. We expand $u_1$ in the $r\to 0$ limit
\begin{equation}
u_1=\frac{i^{n+1}}{\sqrt{2}(\sqrt{2}Q)^{\nu_k}}r^{\nu_k}(1+\cdots).
\end{equation}
Similarly, we can expand $u_2$ and the solution of $\eta_-^{(0)}$. The normalization constants $v_\pm$ are
\begin{equation}
v_\pm=\frac{i^{n+1}}{\sqrt{2}(\sqrt{2}Q)^{\nu_k}}\begin{pmatrix}
\pm1\\
1
\end{pmatrix}.
\end{equation}
The asymptotic behavior near the boundary is
\begin{equation}
\eta_\pm^{(n)}\to a_\pm^{(n)}r^m\begin{pmatrix}
1\\
0
\end{pmatrix}+
b_\pm^{(n)}r^{-m}\begin{pmatrix}
0\\
1
\end{pmatrix},\qquad r\to\infty.
\end{equation}
Consequently, the Green's function near $\omega=0$ to the first order is \cite{Faulkner:2009wj}
\begin{equation}
G_R(\omega,k)=\frac{b_+^{(0)}+\omega b_+^{(1)}+{\cal G}_k(\omega)\bigl(b_-^{(0)}+\omega b_-^{(1)}\bigr)}{a_+^{(0)}+\omega a_+^{(1)}+{\cal G}_k(\omega)\bigl(a_-^{(0)}+\omega a_-^{(1)}\bigr)}.
\end{equation}

We only summarize the result of the perturbation method given in Appendix C of Ref.~\cite{Faulkner:2009wj}. Some notations are slightly changed here. Define
\begin{align}
J^t &=(\bar{\Psi}_0,\Gamma^t\Psi_0)=-\int_0^\infty dr\sqrt{g_{rr}(-g^{tt})}\,(\eta_+^{(0)})^\dagger\eta_+^{(0)}\nonumber\\
J^x &=(\bar{\Psi}_0,\Gamma^x\Psi_0)=\int_0^\infty dr\sqrt{g_{rr}g^{xx}}\,(\eta_+^{(0)})^\dagger\sigma_3\eta_+^{(0)},\label{eq:Jtint}
\end{align}
which are integrations of hypergeometric functions in our system. Note that both $J^t$ and $J^x$ are negative. The various functions in the Green's function Eq.~(\ref{eq:green}) are determined as follows:
\begin{equation}
v_F=\frac{J^x}{J^t},\qquad Z=-\frac{(b_+^{(0)})^2}{J^t},\qquad h=-\frac{v_-^\dagger i\sigma^2v_+}{J^t},
\end{equation}
where the quantities above are evaluated at $k=k_F$.

By plugging $u_1$ and $u_2$ into Eq.~\eqref{eq:Jtint}, the integration can be evaluated for non-negative integers $n$. The first three results of $J^t$ are
\begin{align}
J^{t\,(0)} &=-\frac{(4\nu-1)\sqrt{\pi/2}\,\Gamma(\nu-1/2)}{8Q\,\Gamma(\nu+1)}\nonumber\\
J^{t\,(1)} &=-\frac{(8\nu^2+6\nu-1)\sqrt{\pi/2}\,\Gamma(\nu-1/2)}{8(2\nu+1)^2Q\,\Gamma(\nu+1)}\nonumber\\
J^{t\,(2)} &=-\frac{(8\nu^2+10\nu-1)\sqrt{\pi/2}\,\Gamma(\nu-1/2)}{8(2\nu+1)^2Q\,\Gamma(\nu+2)}.
\end{align}
By induction, we find that the $n$th $J^t$ is given by
\begin{align}
J&^{t\,(n)} = \nonumber \\
 &-\frac{n!\sqrt{\pi}[8\nu^2+(4n+2)\nu-1]\Gamma(\nu+1/2)\Gamma(\nu-1/2)}
{2^{n+3}\sqrt{2}Q(2\nu+1)\Gamma(\nu+n/2+1/2)\Gamma(\nu+n/2+1)},
\end{align}
where $\nu=\nu_k^{(n)}$.

\begin{figure}
\includegraphics[]{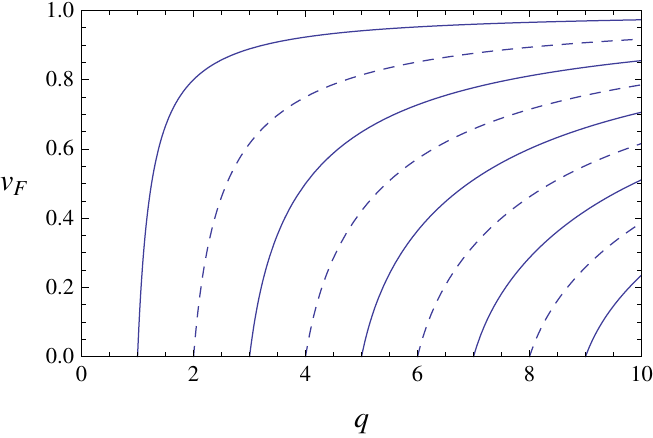}
\caption{\label{fig:vFq} Fermi velocity as a function of charge $q$, where $q>n+1$ for the $n$th Fermi surface. The solid lines are for $n=0$, $2$, $\cdots$, and the dashed lines are for $n=1$, $3$, $\cdots$ (from left to right).}
\end{figure}

After we obtain $J_x$, the Fermi velocity is
\begin{equation}
v_F^{(n)}=\frac{2(2\nu+1)(2\nu-1)}{8\nu^2+(4n+2)\nu-1},
\end{equation}
where $\nu=\nu_k^{(n)}$. When $n$ is even, $v_F^{(n)}$ is for $G_1$; when $n$ is odd, $v_F^{(n)}$ is for $G_2$. If we take $\sqrt{2}Q=1$, the only independent parameter is the charge of the spinor, in terms of which the $v_F$ can be written as
\begin{equation}
v_F^{(n)}=\frac{4(q-n)(q-n-1)}{4q^2-3(2n+1)q+2n(n+1)}.
\end{equation}
We can see that $0\le v_F<1$, and $v_F\to 1$ as $q\to\infty$. The Fermi velocities as a function of the charge is plotted in Fig.~\ref{fig:vFq}.

\begin{figure}
\includegraphics[]{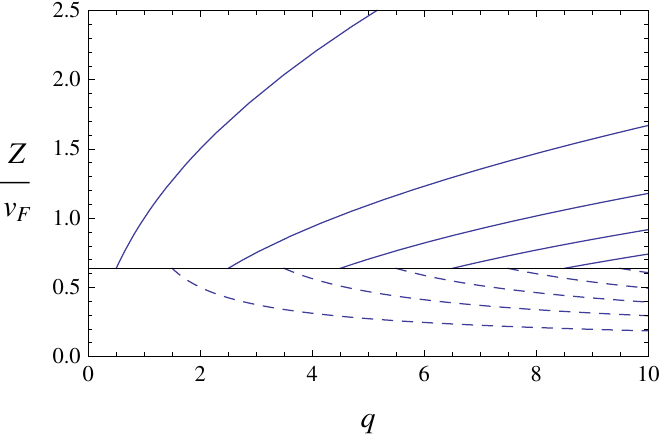}
\caption{\label{fig:ZvFq} $Z/v_F$ as a function of charge $q$, where $q>n+1/2$ for the $n$th Fermi surface. When $q=n+1/2$, $Z/v_F=2/\pi$, as indicated by the horizonal line. The solid lines are for $n=0$, $2$, $\cdots$, and the dashed lines are for $n=1$, $3$, $\cdots$ (from left to right).}
\end{figure}

\begin{figure}
\includegraphics[]{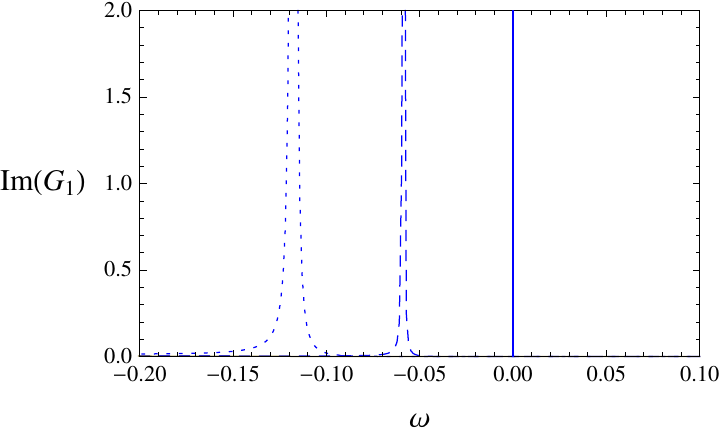}
\caption{\label{fig:rho1} Spectral density as a function of $\omega$ at three different values of $k$. We plot the near $\omega=0$ region for $q=4.5$ and $n=2$, which give $k_F=2$. The dotted, dashed, and solid curves are for $k=1.8$, $1.9$, and $2$, respectively. When $k=k_F$, the quasiparticle peak becomes a pole.}
\end{figure}

If $n$ is even, for the Green's function $G_1$,
\begin{equation}
Z^{(n)}=\frac{2\sqrt{2}Q\Gamma(n/2+1/2)\Gamma(\nu+n/2+1)}
{\pi\Gamma(n/2+1)\Gamma(\nu+n/2+1/2)}v_F^{(n)};\label{eq:Zeven}
\end{equation}
if $n$ is odd, for the Green's function $G_2$,
\begin{equation}
Z^{(n)}=\frac{2\sqrt{2}Q\Gamma(n/2+1)\Gamma(\nu+n/2+1/2)}
{\pi\Gamma(n/2+1/2)\Gamma(\nu+n/2+1)}v_F^{(n)},\label{eq:Zodd}
\end{equation}
where $\nu=\nu_k^{(n)}$. The ratio $Z/v_F$ as a function of charge is plotted in Fig.~\ref{fig:ZvFq}. The self-energy is given by
\begin{equation}
\Sigma^{(n)}=\frac{\Gamma(2\nu+n+1)\Gamma(1/2-\nu)e^{i\pi(1/2-\nu)}\omega^{2\nu}}
{2^{6\nu-1}(\sqrt{2}Q)^{2\nu-1}\Gamma(n+1)\Gamma(\nu+1/2)^3}v_F^{(n)},\label{eq:Sigma}
\end{equation}
where $\nu=\nu_k^{(n)}$. The spectral density $\rho=\text{Im}(G)$ as a function of $\omega$ at different values of $k$ is plotted in Fig.~\ref{fig:rho1}. As $k\to k_F$, the quasiparticle peak will become a delta function at $k=k_F$.

Another way to write down the Green's function is
\begin{equation}
G(\omega,k)=\frac{h_1}{k_\perp-\frac{1}{v_F}\omega-h_2e^{i\gamma_{k_F}}\omega^{2\nu_{k_F}}},\label{eq:Gh}
\end{equation}
where $k_\perp=k-k_F$, $h_1=Z/v_F$, $h_2=|\Sigma/(\omega^{2\nu}v_F)|$, and
\begin{equation}
\gamma_k=\pi(1/2-\nu_k)+\arg\Gamma(1/2-\nu_k).
\end{equation}
The Green's function in the form of Eq.~\eqref{eq:Gh} is analyzed in Ref.~\cite{Faulkner:2009wj} in detail. The poles never appear in the upper half complex $\omega$-plane of the physical sheet. The three cases $\nu_k>1/2$, $\nu_k=1/2$, and $\nu_k<1/2$ correspond to Fermi liquid, marginal Fermi liquid, and non-Fermi liquid, respectively.

The Green's function for the non-Fermi liquid ($\nu_k<1/2$) can be written as
\begin{equation}
G=-\frac{c_1k_\perp^{1/2\nu_k-1}}{\omega-c_2k_\perp^{1/2\nu_k}},
\end{equation}
where
\begin{equation}
c_1=\frac{h_1}{2\nu(h_2e^{i\gamma_k})^{1/2\nu_k}},\qquad c_2=\frac{1}{(h_2e^{i\gamma_k})^{1/2\nu_k}}.
\end{equation}
The residue vanishes as $k\to k_F$. The pole $\omega_*$ moves along a line approaching the origin with the angle
\begin{equation}
\theta_*=\arg(\omega_*)=\begin{cases}
\bigl(\frac{1}{2}+\frac{1}{4\nu_k}\bigr)\pi\quad & k<k_F\\
\bigl(\frac{1}{2}-\frac{1}{4\nu_k}\bigr)\pi\quad & k>k_F.
\end{cases}
\end{equation}
We can see that $\theta_*\notin(0,\pi)$, which is the upper half plane of the physical sheet $\theta\in(-\pi/2,3\pi/2)$. There is a particle-hole symmetry due to $\text{Im}(\omega_*)|_{k_\perp}=\text{Im}(\omega_*)|_{-k_\perp}$.

What is especially interesting is the marginal Fermi liquid. In the $\nu_k\to (1/2)^+$ limit,
\begin{equation}
\Sigma^{(n)}=-\omega-\omega(2\ln\omega-i\pi+\tilde{c}^{(n)})\epsilon+\mathcal{O}(\epsilon^2),
\end{equation}
where $\epsilon=\nu_k-1/2$, and $\tilde{c}^{(n)}$ is a real constant. We can see that the $-\omega$ in $\Sigma$ exactly cancels the $-\omega$ in the denominator of Eq.~\eqref{eq:green}, which is a delicate cancellation between the UV and the IR data. The Green's function for the marginal Fermi liquid is
\begin{equation}
G=\frac{h_1}{k_\perp+\frac{1}{2}(n+1)\omega\ln\omega+c^{(n)}\omega},
\end{equation}
where $h_1$ and $c^{(n)}$ can be easily obtained by the exact solution. The residue also vanishes as $k\to k_F$.

\section{Green's function at arbitrary $\omega$}\label{sec:full}
\begin{figure}
\includegraphics[]{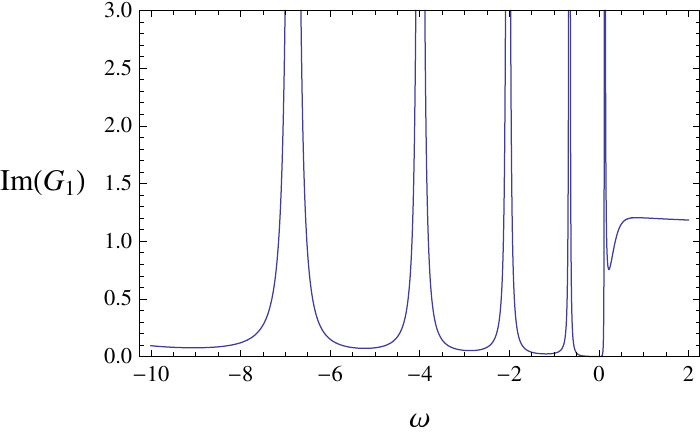}
\caption{\label{fig:rho2} Spectral density as a function of $\omega$ at $k=2$ with $q=10$ and $\sqrt{2}Q=1$. We can see five peaks for the quasibound states. As we increase $k$ from small $k$, the peaks will move to the right. Each time a pole go across $\omega=0$, we obtain a Fermi momentum. Therefore, there are five Fermi surfaces from $G_1$, as the formula of $k_F^{(n)}$ predicts.}
\end{figure}

To obtain the Green's function when $\omega$ is not small, we can solve the Dirac equation numerically with the boundary condition near the horizon as Eqs.~\eqref{eq:uIplus} and \eqref{eq:uIminus}. Alternatively, we will solve the flow equation for $\xi=u_2/u_1$ as follows
\begin{multline}
\partial_r\xi=-\frac{2m}{\sqrt{g^{rr}}}\xi
+\Bigl(\sqrt{\tfrac{|g^{tt}|}{g^{rr}}}(\omega+qA_t)+\sqrt{\tfrac{g^{xx}}{g^{rr}}}k\Bigr)\\
+\Bigl(\sqrt{\tfrac{|g^{tt}|}{g^{rr}}}(\omega+qA_t)-\sqrt{\tfrac{g^{xx}}{g^{rr}}}k\Bigr)\xi^2,
\end{multline}
with the boundary condition $\xi|_{r=0}=i$ ($\omega\neq 0$). The Green's function is obtained by $r^{2m}\xi|_{r\to\infty}$. A typical case of the spectral density $\rho=\text{Im}(G)$ as a function of $\omega$ is plotted in Fig.~\ref{fig:rho2}, in which the peaks correspond to quasibound states.

\begin{figure}
\includegraphics[]{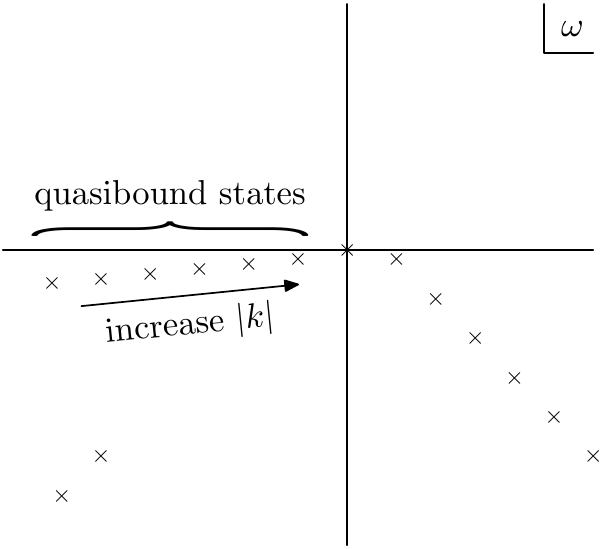}
\caption{\label{fig:qnm} Schematic plot of the poles of the Green's function. The generic feature is that there are quasibound states when the charge of the spinor is sufficiently large. The highly damped modes are plotted from the RN-AdS black hole system.}
\end{figure}

The non-analytic features of the Green's function from the RN-AdS black hole at finite temperature are studied in detail in Ref.~\cite{Herzog:2012kx}. The poles of the Green's function are schematically plotted in Fig.~\ref{fig:qnm}, in which we ignore a small difference that the poles in the finite temperature case cannot be exactly at the origin $\omega=0$. The RN-AdS black hole system and our system have some similar features, as follows. Consider the $m=0$ case. There are no poles in the Green's function at $k=0$, so we start with a small $k$. If the charge of the spinor $q$ is sufficiently large, there are quasibound states. As we increase $k$, the poles with $\text{Re}(\omega)<0$ will move to the right. When a pole goes through the origin $\omega=0$, we obtain a normal mode, which indicates a Fermi surface. The number of the quasibound states equals the number of the Fermi surfaces.

\begin{figure}
\includegraphics[]{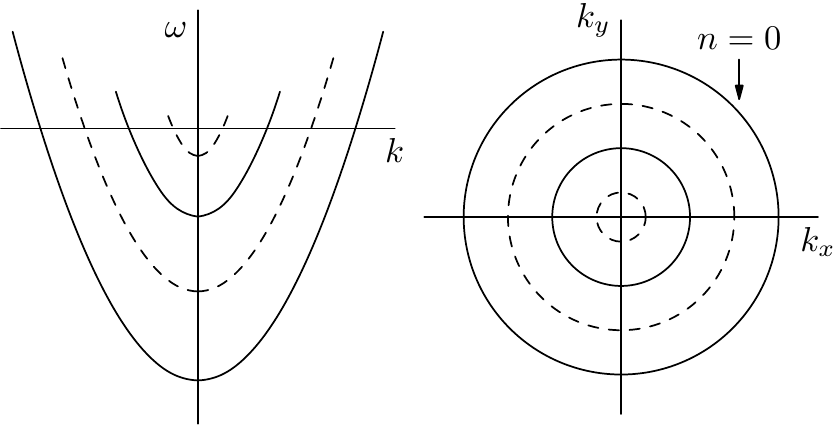}
\caption{\label{fig:FS} Schematic plot of dispersion relation and the Fermi surface by the massless spinor in the bulk (for massive spinor in the bulk, see Ref.~\cite{Herzog:2012kx}). The system has rotational invariance, and we only show some intersections. If we decrease the charge $q$, the Fermi surfaces will shrink but keep the same space.}
\end{figure}

The schematic plots of the dispersion relation and the Fermi surface are shown in Fig.~\ref{fig:FS}. Our system has at least one Fermi surface when $q>1/2$. If we increase $q$, more Fermi surfaces will appear, and the Fermi surfaces are equally spaced.


\begin{figure*}
\begin{minipage}[t]{\textwidth}
\includegraphics[]{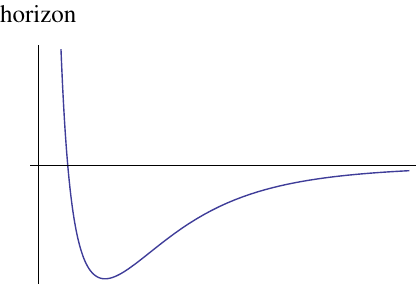}\qquad
\includegraphics[]{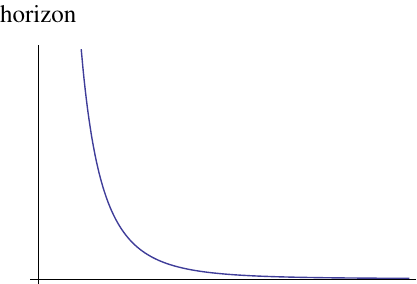}
\caption{\label{fig:Veff0} Effective potential when $\omega=0$.}
\end{minipage}\\[15pt]
\begin{minipage}[t]{\textwidth}
\includegraphics[]{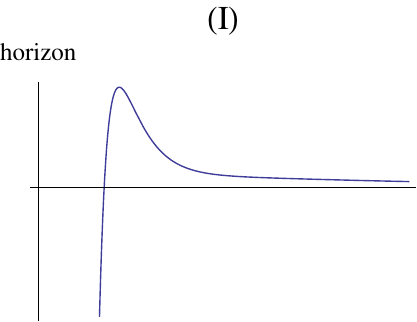}\quad
\includegraphics[]{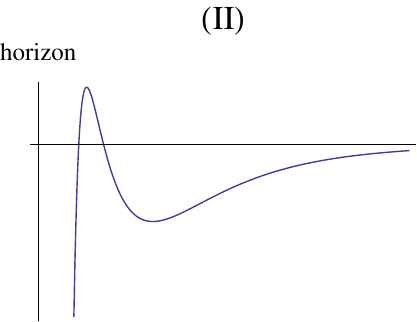}\quad
\includegraphics[]{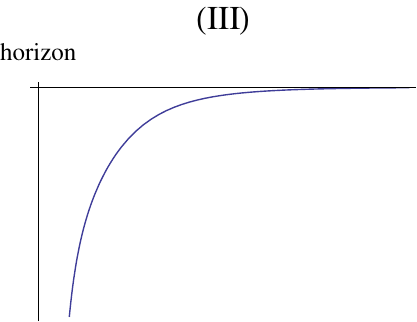}
\caption{\label{fig:Veff1} Effective potential when $\omega\neq 0$.}
\end{minipage}
\end{figure*}

The number of quasibound states can be estimated by the WKB method. The effective potential is\footnote{This is the leading order of the effective potential, in the sense of Ref.~\cite{Hartnoll:2011dm}. Higher order terms contain singularities. A more rigorous WKB treatment is in Ref.~\cite{Herzog:2012kx}, which shows that the singularities in higher order terms are essential to the negative sign of the imaginary part of the quasinormal modes.}
\begin{equation}
V_{\rm eff}=\frac{m^2}{g^{rr}}+\frac{g^{xx}}{g^{rr}}k^2-\frac{|g^{tt}|}{g^{rr}}(\omega+q\Phi)^2.
\end{equation}
The distinctive shapes of the effective potential are plotted in Figs.~\ref{fig:Veff0} and \ref{fig:Veff1}. At the AdS boundary, $V_{\rm eff}=0$. The near horizon behavior is
\begin{multline}
V_{\rm eff}\to
-\frac{\omega^2}{4r^4}+\left(\frac{k^2}{2Q^2}-\frac{q\omega}{\sqrt{2}Q}-\frac{\omega^2}{4Q^2}\right)\frac{1}{r^2}\\
+\frac{m^2}{2Q^{2/3}r^{4/3}}+\cdots.
\end{multline}
We need to treat $\omega=0$ and $\omega\neq 0$ cases separately. Assume at least one of $k$ and $m$ is nonzero, otherwise the Green's function has no poles.

When $\omega=0$, the leading term in $V_{\rm eff}$ is positive as $r\to 0$, which implies that the state cannot tunnel to the horizon and thus is stable. This is similar to the electron star, but different from the extremal RN-AdS black hole, in which there are no stable bound states when $\nu_k$ is imaginary.

When $\omega\neq 0$, the leading term in $V_{\rm eff}$ is negative, which implies that there are no exact bound states with $\omega\neq 0$. The state can tunnel through a barrier to the horizon, which will lead to an imaginary part of the modes. The qualitative features of the effective potential are similar to the RN-AdS black hole. If $q$ is large enough, there is a potential well with a barrier. The quasibound states in the well can tunnel through the barrier. In Fig.~\ref{fig:qnm}, the modes near the real $\omega$ axis correspond to the quasibound states.

\section{Discussion}\label{sec:sum}
Starting from a dilatonic black hole derived from a consistent truncation of type IIB supergravity, we have studied the fermionic Green's function dual to massless fermions in the bulk.  We obtained exact analytic results at zero frequency, and exact asymptotic results for small frequencies.  These analytic results capture key features of the strongly coupled fermionic system modeled by the gauge/gravity duality, and they provide a new universality class for the strange metal phase at quantum criticality. Provided that the charge of the bulk fermion is not too small, there are Fermi surfaces. Their Fermi momenta are equally spaced, and there are a finite number of them, approximately proportional to the charge of the bulk fermion. The IR scaling dimension is always real. The properties of the RN-AdS black hole system and our system are compared and summarized in Table~\ref{tab:sum}.

\begin{table*}
\caption{\label{tab:sum} Comparison between the RN black hole system and our system, with the same UV geometry as AdS$_5$.}
\begin{ruledtabular}
\begin{tabular}{lll}
 & RN black hole & two-charge black hole\\\hline
Charge & $Q_1=Q_2=Q_3=Q$ & $Q_1=Q_2=Q$, $Q_3=0$\\
Entropy & $S=\text{constant}$ & $S\propto T\to 0$\\
IR (near horizon) geometry & $\text{AdS}_2\times\mathbb{R}^3$ & conformal to $\text{AdS}_2\times\mathbb{R}^3$\\
Electric field near horizon & $E=\text{constant}$ & $E\propto r\to 0$\\
Stability near horizon & unstable due to pair production & stable against pair production\\
IR scaling exponent $\nu_k$ & $\nu_k\propto\sqrt{k^2-k_o^2}$ & $\nu_k\propto k$\\\cline{2-3}
$\nu_k\gtreqqless 1/2$ & \multicolumn{2}{c}{Fermi liquid, marginal Fermi liquid, non-Fermi liquid}\\
\end{tabular}
\end{ruledtabular}
\end{table*}

There are several instabilities that can modify the bosonic background, including the superconducting and the Gregory-Laflamme instabilities; however, these instabilities all involve extra fields not present in our consistent truncation of the supergravity Lagrangian, Eq.~\eqref{eq:StartingLagrangian}.

\acknowledgements
J.R. thanks Prof.~C.P.~Herzog for his guidance on the topics of holography and condensed matter physics, and thanks Prof.~A.M.~Polyakov for helpful discussions. This work was supported in part by the Department of Energy under Grant No.~DE-FG02-91ER40671.

\appendix
\section{Mathematical notes}\label{sec:notes}
\textit{Hypergeometric function ${_2F_1}(\alpha,\beta;\gamma;z)$.} We will denote ${_2F_1}$ by $F$ for simplicity in the following. The derivative of the hypergeometric function can also be expressed by a hypergeometric function:
\begin{equation}
\frac{d}{dz}F(\alpha,\beta;\gamma;z)=\frac{\alpha\beta}{\gamma}F(\alpha+1,\beta+1;\gamma+1;z).
\end{equation}
The following formula can be used to combine the sum of two hypergeometric functions:
\begin{multline}
\gamma F(\alpha,\beta;\gamma;z)+\alpha zF(\alpha+1,\beta+1;\gamma+1;z)\\
=\gamma F(\alpha,\beta+1;\gamma;z).
\end{multline}
In general, the hypergeometric function has branch points at $z=0$, $1$, and $\infty$. By convention, we make a branch cut from $z=1$ to $\infty$, and take the principle branch as $-2\pi<\arg z\leq 0$ for $|z|>1$. The following formula can be used to transform a value above the branch cut to another value below the branch cut:
\begin{equation}
F(\alpha,\beta;\gamma;z)=(1-z)^{-\alpha}F\bigl(\alpha,\gamma-\beta;\gamma;\frac{z}{z-1}\bigr).\label{eq:trans}
\end{equation}
The $z=2$ point has the following special property:
\begin{equation}
z\to\frac{z}{z-1}:\qquad 2\pm i\epsilon\to 2\mp i\epsilon.
\end{equation}
By Eq.~\eqref{eq:trans} and $(-1-i\epsilon)^{-\alpha}=e^{i\alpha\pi}$, we have
\begin{equation}
F(\alpha,\beta;\gamma;2+i\epsilon)=e^{i\alpha\pi}F(\alpha,\gamma-\beta;\gamma;2-i\epsilon).
\end{equation}

We define
\begin{equation}
F(\alpha,\beta;\gamma;2):=F(\alpha,\beta;\gamma;2-i\epsilon).
\end{equation}
The condition for the normal modes is
\begin{equation}
F(-n,\nu_k;2\nu_k+1;2)=\pm F(-n,\nu_k+1;2\nu_k+1;2), \label{eq:nmodes}
\end{equation}
where $n=-\nu_k+q-1/2$. The plus sign is for $G_1$ and the minus sign is for $G_2$. We assume that $n$ is a non-negative integer at first. If $n$ is even, the above equation with the plus sign is satisfied; If $n$ is odd, the above equation with the minus sign is satisfied. We can numerically check that they are the only solutions when $q>0$. When $q<0$, there is another set of solutions due to the $q\to -q$, $\omega\to -\omega$, $u_1\leftrightarrow u_2$ symmetry of the Dirac equation; however, these solutions are unphysical because they give $\text{Im}(G)<0$. Intuitively, only if a particle and the black hole have the same charge can there be a balance between the attractive gravitational force and the repulsive electromagnetic force on the particle.

For non-negtive integer $n$,
\begin{multline}
F(-n,\nu;2\nu+1;2)\\
=\begin{cases}
\dfrac{\Gamma(n/2+1/2)\Gamma(\nu+1/2)}{\sqrt{\pi}\,\Gamma(\nu+n/2+1/2)}\, &\text{if $n$ is even}\\
\dfrac{\Gamma(n/2+1)\Gamma(\nu+1/2)}{\sqrt{\pi}\,\Gamma(\nu+n/2+1)}\, &\text{if $n$ is odd.}
\end{cases}
\end{multline}

\textit{Gamma function.} Useful identities for the Gamma functions include
\begin{gather}
\Gamma\Bigl(n+\frac{1}{2}\Bigr)=\frac{(2n)!}{4^nn!}\sqrt{\pi}=\frac{(2n-1)!!}{2^n}\sqrt{\pi}\nonumber\\
\Gamma(z)\Gamma(1-z)=\frac{\pi}{\sin\pi z}\nonumber\\
\Gamma(z)\Gamma(z+1/2)=2^{1-2z}\sqrt{\pi}\,\Gamma(2z).
\end{gather}

\textit{Whittaker function.} Whittaker's equation is
\begin{equation}
\frac{d^2W}{dz^2}+\left(-\frac{1}{4}+\frac{\lambda}{z}+\frac{1/4-\mu^2}{z^2}\right)W=0.
\end{equation}
We can write down the general solution as
$C_1W_{\lambda,\mu}(z)+C_2W_{-\lambda,\mu}(-z)$, where for large $|z|$ one has
\begin{equation}
W_{\lambda,\mu}(z)\sim e^{-z/2}z^\lambda(1+\cdots),\qquad |z|\to\infty.
\end{equation}
As special cases, $W_{\pm 1/2,\mu}(z)$ are related to the modified Bessel function $K_\nu(z)$ by
\begin{align}
W_{1/2,\mu}(z) &=\frac{z}{2\sqrt{\pi}}\left(K_{\mu+1/2}\left(\frac{z}{2}\right)+K_{\mu-1/2}\left(\frac{z}{2}\right)\right)\nonumber\\
W_{-1/2,\mu}(z) &=\frac{z}{2\mu\sqrt{\pi}}\left(K_{\mu+1/2}\left(\frac{z}{2}\right)-K_{\mu-1/2}\left(\frac{z}{2}\right)\right).
\end{align}


\onecolumngrid
\vspace{5mm}
\twocolumngrid

\end{document}